# Bitcoin With Heterogeneous Block Sizes: A Scaling Proposal

Santi J. Vives

jotasapiens.com/research

**Abstract:** We propose a bitcoin generalization as a solution to the problem of scalability. The block is redefined as a sequence of sub-blocks of increasing sizes that coexist as different levels of compromise between decentralization and transactions throughput. Miners and users can decide individually the size they use without affecting others.

## 1. Introduction

Bitcoin is the first mathematical (cryptographic) construction to enable secure, permissionless, and peer-to-peer money and payments. It solves many fundamental problems, but others still need work. As it gains adoption, bitcoin begins to face problems of scalability. The volume of transactions the network is capable of handling remains small compared to traditional payment systems. The limit of transactions is currently determined by a maximum block size agreed by the network. The block size can be increased over time as hardware becomes more efficient. But increasing the size too much or too early can lead to a centralized network, with a few miners controlling most of the hashrate.

The problem can be described in a simple way. In the bitcoin network, we find a set of different miners, with different hardware, capable of processing different volumes of transactions. We can sort miners by the maximum block size they can handle. From the smallest to the largest, we get:

$m_0 < m_1 < m_2 < m_3 < ... < m_N$

It's easy to see that if choose the size the smallest miner $m_0$ can handle, every other miner will be able to process the block as well, resulting in the most decentralization. On the opposite side, if we choose the largest size $m_N$ can handle, the entire network would be controlled by one single miner. It follows that any block size parameter we might choose sets a trade-off between the security and the capacity of the network.

As we move to smaller block sizes, the space available for transactions shrinks and more persons become unable to make payments. The demand combined with a lack of space will raise fees. As fees get higher than the amount of coins stored in many addresses, more persons loose access to their coins. On the other hand, as we move to larger block sizes, the network becomes more centralized, increasing the risk of a miner controlling a majority of the network, censoring transactions, and reversing them.

When they decide a size which implies a trade-off between decentralization and capacity, developers and miners choose between two properties that are the two desirable and important. Since different persons have different preferences between the two, no matter which block size is chosen it will always leave someone unsatisfied, leading to fragmentation and conflict.

Through the paper we will make a series of modifications to the construction, until we arrive to a generalization that allows increasing blocks to arbitrarily large sizes without destroying decentralization. Instead of imposing a single size, multiple sizes coexist as different trade-offs between security and capacity. Miners and users can choose individually the size they use without affecting others.

## 2. Blocks with many sizes

To create the generalized construction, we will start by discarding the idea of having a single, fixed block size in the network. Instead, we will define an infinite sequence of sizes that will coexist in the chain. Each of the sizes corresponds to a different level of compromise between decentralization, and transactions capacity.

To achieve that, we redefine the block as a series of sub-blocks. We start by defining a size for the first sub-block (for example, 1 MB). Each next sub-block doubles the total size, resulting in a sequence of sub-blocks with relative sizes:
1, 1, 2, 4, 8, 16, 32, 64, 128, 256...

Each miner chooses a cutoff point in the series, according to its capabilities. For example, a miner only capable of processing 2 MB blocks, will process only the first sub-blocks with relative sizes *1, 1*. While a miner able to handle 8 MB blocks can process the first four sub-blocks with relative sizes *1, 1, 2, 4*, since they add up to 8.

Since the cutoff can differ among miners, the chain now consists of blocks of different sizes. When large blocks are inserted in the chain, small-block miners are only able to process the first sub-blocks up to their cutoff. For example, if a 8 MB block is generated, the small-block miner with cutoff 2 will only process the first 2 sub-blocks, and ignore the rest.

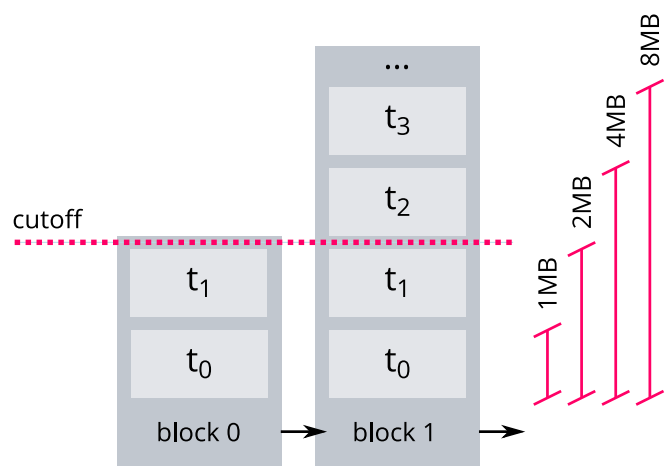

## 3. Heights

We will call the *height* of a sub-block the position it has within the block. The first, smallest sub-block has height 0, the next sub-block has height 1, and so on. Height 0 is processed by every miner in the network. And as we move up, each next height in the series is processed by a smaller number of miners: a height *y* is processed by the set of miners with cutoff ≥ *y*.

Since sub-blocks at different heights are observed and verified by different sets of miners, we will make addresses and transactions at each height independent from the others:

- Users generate addresses that belong to a specific height. Each *0-address, 1-address, 2-address … n-address* corresponds to sub-blocks at height *0, 1, 2 … n*. The height of an address is the parameter that defines the trade-off between decentralization, and transaction capacity.
- Transactions move coins from addresses at one height to other addresses at the same height.
- To generate a new block, miners collect transactions of heights up to their cutoff. They combine transactions that belong in a same height, and combine them in a Merkle tree to form each sub-block.
- As in the original bitcoin construction, miners collect fees and a predefined number of new coins as a reward for their work. In this case, all fees collected from the transactions at different heights, and the newly mined coins are paid to an address at the top. This property will come handy later.

## 4. A stream of sub-blocks

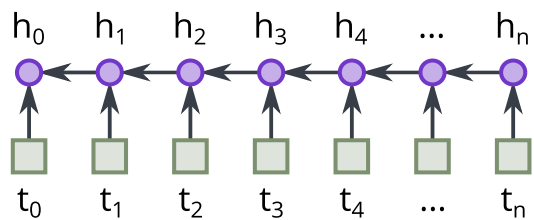

The chain now contains blocks of different sizes, including blocks with sizes that exceed what small-block miners can process. We need a method that allows big-block miners to broadcast only the first sub-blocks to small-block miners, and still allows small-block miners to authenticate them. Small-block miners must be able to verify that the sub-blocks are members of the block, despite not knowing the block entirely. For that we will use a stream, a type of hash tree we will call by that name for its ability to process and authenticate chunks of data of arbitrary length in sequential order.

Every time a block is created, each miner starts with a list of sub-blocks of different heights, each consisting of a Merkle trees with transaction. Let's name the trees, from the smallest to the largest, $t_0$, $t_1$, $t_2$ ... $t_n$. Rather than including the Merkle trees directly, each block will include a root of a stream tree, with one of the Merkle tree at each of its leaves. To compute the stream, the miner hashes the last tree $t_n$ to obtain a value $h_n$. Then the miner hashes the previous tree $t_{n-1}$ together with $h_n$ to obtain a new value $h_{n-1}$. The process is repeated from the last tree to the first, until obtaining the root of the stream $h_0$. Generally, a stream with root $h_0$ that contains $n+1$ elements $t_0...t_n$ is defined as:

$$h_y = \begin{cases} hash(t_y, h_{y+1}) & if\ y < n \\ hash(t_n) & if\ y == n \end{cases}$$

A sequence of trees from the stream can be authenticated with a single hash value. To prove a sequence of $t_0...t_y$ we only need $h_{y+1}$. By hashing them in the same order the stream was created, we can authenticate them against the root $h_0$. For example, a big-block miner with cutoff *6* will produce a block with trees $t_0...t_6$. When connecting to a miner with cutoff *2*, the big-block miner needs to send trees $t_0...t_2$, and hash value $h_3$.

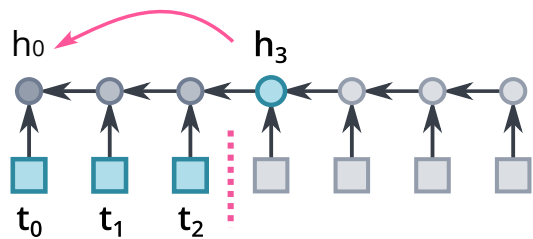

Miners can decrease or increase their cutoff at any time. In our example, the small-block miner could lower the cutoff from *2* to *1* by dropping every tree $t_2$, and storing $h_2$. Instead, to increase the cutoff from *2* to *4*, the miner needs to download trees $t_3...t_4$, and value $h_5$. The miner can authenticate appended trees and $h$ value by hashing them, and comparing them against $h_3$.

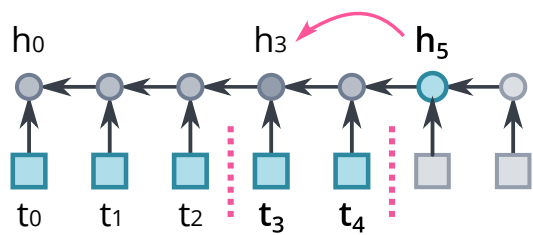

## 5. Separation of extension and confirmation

Each miner can now choose a size cutoff, and process part of the blocks without knowing what's happening above. Still, big blocks need not to affect the security of small blocks.

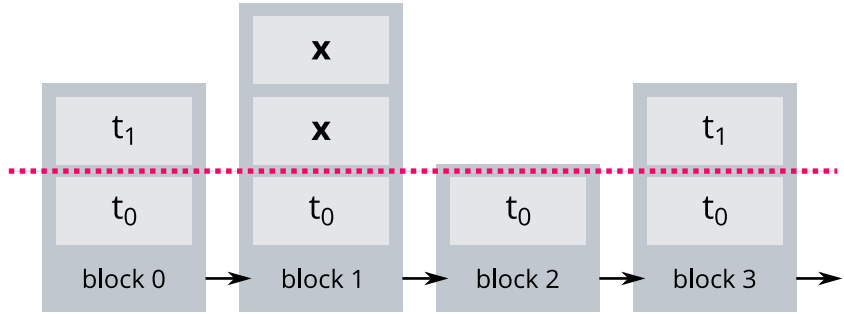

Let's consider an example. There are 2 groups of miners, one processing small blocks, and the other bigger ones. Among the big-block group, one dishonest miner publishes a block that contains valid transactions at the lowest height, but invalid transactions at the heights above. Next, an honest small-block miner, who sees the block as valid, extends the chain with a valid small block. A third honest big-block miner is working to extend the chain. If the third miner rejects blocks containing any invalid information, he would have to reject the last two blocks. Then, the honest small-block miner worked on a chain that seemed valid, only to be rejected later.

It follows that if we want to work on the chain without observing all information, and we don't want bigger blocks to affect the security of smaller ones, extending the chain cannot be considered as confirmation of past information.

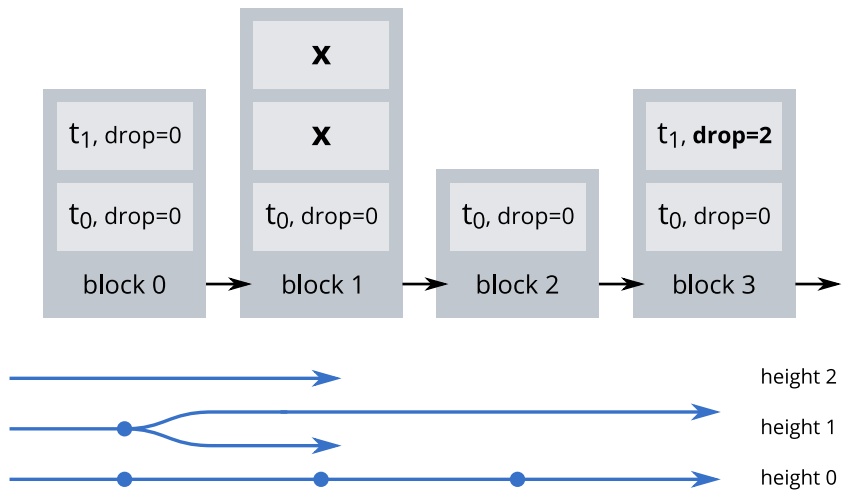

To solve the problem, let's separate confirmation from extension. At each height we will include the number of previous sub-blocks from that same height that we drop, either because they are invalid or because they don't exist. A sub-block at position *(x, y)* that drops *n* previous sub-blocks confirms sub-block *(x−n−1, y)*. In addition, every time we drop a sub-blocks, we will also drop all sub-blocks in the heights above. This is to ensure no new coins are mined, and no fees are paid when sub-blocks are rejected.

Similarly to the original bitcoin construction, miners work to extend the longest valid chains of sub-blocks at each height. If two of the same length are valid, they extend the one that appears first on the main chain.

## 6. Moving up

At this point the coins stored at different heights are completely independent. We could exchange them as we would do with different coins, each with its price. But it would be better to define a method to move coins between heights.

We'll start by moving a coin to an upper height, which is easier. Suppose we own a coin at height *y*, and we want to move it to an address in an upper height *y+n* (for example, from a 0-address to a 1-address). We need to insert the transaction at height *y*. Once the transaction is included in a block:

- Miners with cutoff less than *y* ignore the transaction.
- Miners with cutoff at least *y* (but less than *y+n*) verify the signature, and remove the coin from their database.
- Miners with cutoff *y+n* or higher verify the signature, remove the coin from the database at the lower height, and include it in the receiving address at the upper height.

## 7. Moving down

Now we need to figure out how to move coins down. Suppose we own one unspent coin at height *y*, and we want to move it to the height below, *y−1*. To start, we need to broadcast a transaction to be inserted at height *y*. Big-block miners with cutoff *y* or higher can verify we own an unspent coin, and that we want to move the coin from the y-address to a *(y−1)*-address. But a problem arises: small-block miners with cutoff *y−1* can't observe the transaction. They can't even verify that we own the coin.

To solve this, we need to make some modifications. Big-block miners can claim that a resulting coin exists at the lower height after moving down, but they can never prove it to small-block miners. We'll start by reflecting this in the construction.

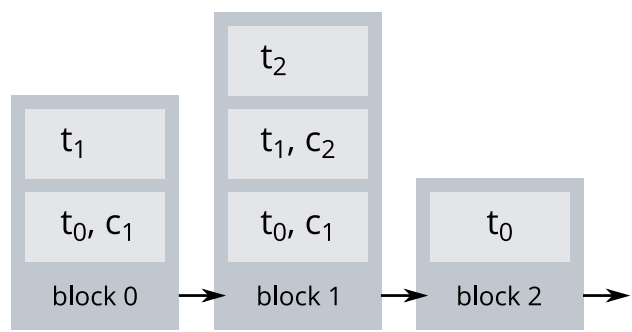

In addition of the tree with regular transaction $t_y$, each sub-block will include a tree with claims $c_y$., as well as distinct drop values for the number of previous transaction trees, and previous claim trees. Big-block miners insert the transaction (moving a coin from *y* to *y−1*) at height *y*, and include a claim that the recipient address has gained one coin at height

*y−1*. When extending the chain, big-block miners must verify that the claims and the transactions match. Miners confirm or reject sub-blocks following these rules:

1. Miners with cutoff *Y* can confirm or reject transactions up to height *Y*, confirm or reject claims up to *Y−1*, and only reject but not confirm claims at height *Y*.
2. Blocks with heights *0…t* are allowed to have claims at heights *0…t−1*.
3. If transactions at some height are rejected, claims at the same height must also be rejected. That is, if regular transactions are invalid the full sub-block is invalid.
4. Transactions at some height can be accepted, even if claims at the same height are rejected.
5. Miners must verify that transactions trees $t_y$ and claim trees $c_{y-1}$ are in agreement, for *y=1…Y*. If either is rejected, the other must be reject as well.
6. If a pair $t_y$, $c_{y-1}$ is rejected, everything above must also be rejected.

## 7.1 Transformation into regular coins

So far, coins that where at an upper height are now observable at the height below, but they remain different and independent from regular coins. They don't even have the same level of security, since they can be reversed if a sub-block above gets rejected. At some point, transfered coins should become the same as regular coins.

Since miners cannot know information above their cutoff, if we want to move coins down irreversibly, at some point we will have to accept them as valid even if the information above is invalid. For that purpose, we will establish a number of confirmations after claims can no longer be reversed. Pass that time, claimed coins will become regular coins. Sub-blocks above will be dropped no matter how old if they include invalid transaction that do not match the claimed coins below, but the claimed coins will remain. Thus, checking that claims and upper transactions are in agreement is a weaker rule, only enforceable for some time.

In upper heights, the hashrate is concentrated among fewer miners, increasing the probability that an attacker could include a false claim without being reversed in the next sub-blocks. To compensate, we will increase the required number of confirmations with height. Thus, a dishonest miner would control a larger portion of the hashrate if he produces larger blocks, but would also need to sustain an attack for a longer time to succeed.

Suppose we lock coins for a basic period of 100 confirmations at height 0, and we increase it linearly with height. A claim from height *y*, as well as fees and mined coins paid to an *y*-address, can be spent after
*(y+1)·100* confirmations

## 7.2 Preventing inflation

Let's consider an example scenario of a miner making a false claim. Suppose there are 100 coins. The dishonest miner includes valid transaction moving 10 coins down from height *y* to height *y−1*. At height *y−1*, the miner claims the 10 coins resulting from the valid transactions, and 1 additional coin to an address of his own not matching any transaction. After the attack, the chain contains 101% the coins, without small-block miners noticing. Clearly, to be secure, the construction needs a mechanisms to prevent dishonest miners from arbitrarily inflating the money supply.

We can solve this problem by introducing a simple rule: a miner can only claim as many coins as would be destroyed if the claim is false.

Let's remember some of the rules we got far. A sub-block is invalid if the transactions do not match the claims below. The invalid sub-block, as well as all sub-blocks at the heights above, will always be rejected and reversed, no matter how old. Since mined coins and transactions fees are paid to an address at the top, those coins will get destroyed. Then, if we want to prevent inflation, we must constraint the amount that can be claimed to the observable amount that would be paid at the top.

For a miner verifying a sub-block at height *y*, it means that the amount claimed must be at most the number of coins mined in that block, plus all the fees collected up to that height, minus the amount already claimed at the heights below:

$$claimed_y \leq mined + \sum_{n=0}^{y} fees_n - \sum_{n=0}^{y-1} claimed_n$$

Miners must verify that all claims up to their cutoff follow this rule. If a sub-block includes a claim tree with an invalid amount, the full sub-block must be rejected. Claims with invalid amounts can never settle, no matter how old.

To sum up, claim validation follows a weak rule and a stronger one:

- Miners must verify that claims are in agreement with the upper transactions they can observe. This rule is weaker and is only enforced for a limited period of time. A sub-block can still be accepted even when its claim is rejected.
- Miners must verify that the amount claimed is at most the amount of fees and mined coins observed up to that height. This is a stronger rule that must be always enforced. If the rule is not satisfied, the entire sub-block must be rejected, no matter how old.

Now, let's revisit the attack with the introduced rule. A dishonest miner can include transactions in disagreement with the claim tree below. The sub-block with transactions will always be rejected by honest miners, no matter how many confirmations. The claim tree can be accepted if it gets enough confirmations, sending coins to addresses of the attacker's choosing. But those coins are at most the same ones the attacker would have gained from mining the block, and collecting fees from the valid transactions below. Thus, the attempted attack now results in the same output as if the invalid pair of transactions and claims had never existed.

## 7.3 Offset

The same principles we've used to move coins one height down can be generalized to move multiple heights. At each sub-block we will include trees of claims moving coins 1, 2, 4, 8... heights down. We will combine regular transactions and all claims in a stream to form each sub-block. Second, we will combine the sub-block using another stream as usual.

Using the notation $c_y^z$ for a tree of claims moving coins from height *z* to *y*, a sub-block at height *y* contains claims of the form $c_y^{y+2^n}$.

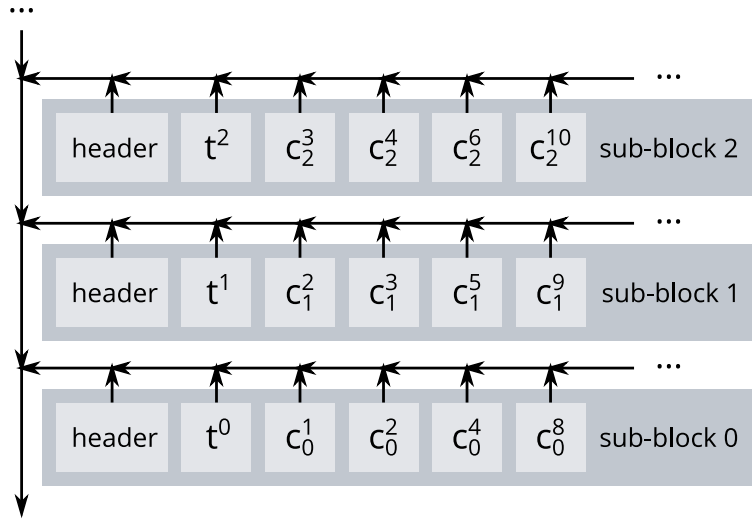

Extending the same principles as before, we get that:

- A sub-block at some height now contains regular transaction and multiple claim trees. Combined, they must be within the allowed size for that height.
- When processing a sub-block, the miner must verify all of the claim trees it contains to ensure the total amount that is claimed is within the allowed limit.
- If a sub-block contains claims from $z$ to $y$, all fees and mined coins must be paid to an address at height at least $z$.
- A claim moving coins from $z$ to any height below must wait the same number of confirmations as previously defined for a claim from $z$ to $z-1$.
- If a claim tree from $z$ to $y$ is dropped, all sub-blocks at heights $z$ and above must be dropped. Then, if a claim tree from $z$ to $y$ is dropped, any following claim tree in the same sub-block (>$z$ to $y$) must be dropped as well, since they move coins from a height that is being rejected.

## 8. Addresses with dynamic height

As time passes, demand to include transactions might increase, while capacity at any height is constant. That increase in demand will result in an increase in transaction fees. For addresses with a small amount of coins, it won't be possible to move funds.

We need to prevent coins from getting stuck due to high fees. For that purpose, let's define a method to move coins to upper heights as times passes. Suppose we want to store a coin, and we estimate that we will use it, at most, within a year. At the time of creating the address we'll define a dynamic height using the following example rules:

- During the first year the coins are spendable by inserting a transaction at height $y$.
- From that point, the coins move 1 height up every 3 months.

Miners with cutoff $y$ can include a transaction spending the coins within the first year, or will remove the address if the year passes. Miners with cutoff $y+1$ will remove the address from their databases after 3 more months. And so on.

## 9. Conclusion

We've learned it is possible to increase bitcoin blocks to arbitrarily large sizes without destroying decentralization, and that scaling can happen on-chain, preserving much of the simplicity in the original construction. The proposed construction redefines a block as a sequence of sub-blocks of increasing sizes that coexist in the same chain as different trade-offs between decentralization and transactions capacity. Each miner, and each user can decide the size to use. By taking away from miners and developers the ability to define a block size for the entire network, the construction takes away the ability to make a decision about other people's coins.

Each miner chooses a cutoff that determines the size of blocks generated by that individual miner, and the number of sub-blocks to process from other miners. Miners can increase or decrease their size cutoff at anytime. The network reaches consensus despite the fact most miners only observe part of the information in the chain.

People can use sub-blocks at lower heights to store their savings, which offer the most security and decentralization at the expense of more waiting time to make payments, and higher fees. While consumers and merchants can use upper heights for regular payment, which are capable of higher transactions throughput and lower fees at the expense of more centralization and an increased risk of transactions reversal. Merchants can compensate it by either waiting for more confirmations, or absorbing the risk as part of the price.